\documentclass[aps,prl,twocolumn,groupedaddress,floatfix,showpacs,longbibliography]{revtex4-1}
\usepackage{graphicx}
\usepackage{amsmath}
\usepackage{color}

\begin{document}

\title{Measuring the Hydrodynamic Linear Response of a Unitary Fermi Gas}

\author{Lorin Baird, Xin Wang, Stetson Roof and J. E. Thomas}

\affiliation{$^{1}$Department of  Physics, North Carolina State University, Raleigh, NC 27695, USA}

\date{\today}

\begin{abstract}
We directly observe the hydrodynamic linear  response of a unitary Fermi gas confined in a box potential and subject to a spatially periodic optical potential that is translated into the cloud at speeds ranging from subsonic to supersonic. We show that the time-dependent change of the density profile is sensitive  to the thermal conductivity, which controls the relaxation rate of the temperature gradients and hence the responses arising from adiabatic and isothermal compression.

\end{abstract}

\maketitle
A unitary Fermi gas is a scale-invariant, strongly interacting quantum many-body system, created by tuning a trapped, two-component cloud near a collisional (Feshbach) resonance~\cite{OHaraScience}. Unitary gases are of great interest~\cite{BlochReview}, as the thermodynamic properties and transport coefficients are universal functions of the density and temperature, enabling parameter-free comparisons with predictions. Equilibrium thermodynamic properties of trapped unitary gases have been well characterized~\cite{ThermoLuo,KuThermo}. In contrast, hydrodynamic transport measurements require dynamical experiments that have been obscured by the low density near the cloud edges, which leads to free streaming. For expanding clouds~\cite{CaoViscosity,JosephShearNearSF}, this problem has been circumvented by employing second order hydrodynamics methods to extract the local shear viscosity~\cite{BluhmSchaeferModIndep,BluhmSchaeferLocalViscosity}, and is obviated for trapped samples with uniform density. A normal unitary gas, at temperatures above the superfluid transition, is a single component fluid that affords the simplest universal system for hydrodynamic transport measurements, as the transport properties comprise only the shear viscosity $\eta$ and the thermal conductivity $\kappa_T$, since the bulk viscosity vanishes in scale-invariant systems~\cite{SonBulkViscosity,StringariBulk,ElliottScaleInv}. Further, measurements in the normal fluid at high temperature $T$ can be compared with benchmark variational calculations for a unitary gas in the two-body Boltzmann limit~\cite{BruunViscousNormalDamping,BrabySchaeferThermalCond},
\begin{equation}
\eta=\frac{15}{32\sqrt{\pi}}\frac{(mk_BT)^{3/2}}{\hbar^2}
\label{eq:etaHighT}
\end{equation}
and
\begin{equation}
\kappa_T=\frac{15}{4}\frac{k_B}{m}\,\eta,
\label{eq:kappaHighT}
\end{equation}
with $k_B$ the Boltzmann constant and $m$ the atom mass.

In this Letter, we demonstrate a new probe of hydrodynamic transport, which is applied to a normal unitary Fermi gas of $^6$Li. The gas is confined in a repulsive box potential, creating a sample of nearly uniform density, and driven by a moving, spatially periodic optical potential of chosen wavelength $\lambda$ along one axis $z$, which moves into the box at a selected speed $v$. We measure the density response $\delta n(z,t)$, which is analyzed using a linear hydrodynamics model. The model shows that the response profiles are sensitive to the effective sound speed, which is controlled by the ratio of the tunable wave frequency $\omega=2\pi v/\lambda$ to the decay rate of the temperature gradients, $\gamma_\kappa\propto\kappa_T/\lambda^2$.  When $\gamma_\kappa<<\omega$, temperature gradients relax slowly and sound waves propagate at the adiabatic sound speed $c_0$. In the opposite limit, $\gamma_\kappa>>\omega$, temperature gradients relax quickly and sound waves propagate at the isothermal sound speed $c_T<c_0$.

\begin{figure}[htb]
\begin{center}\
\includegraphics[width=3.25in]{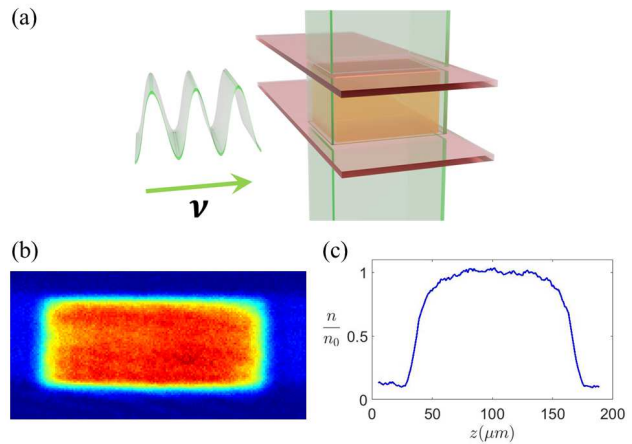}
\end{center}
\caption{A unitary Fermi gas, confined in a box, is driven by a moving spatially periodic potential. (a) The box potential is created by two 669 nm sheet beams (top/bottom) and four vertically propagating 532 nm sheet beams; (b) Column density;  (c) Integrated column density in the box potential showing 1D profile.
\label{fig:boxdensity}}
\end{figure}

The experiments, Fig.~\ref{fig:boxdensity}, employ ultracold $^6$Li atoms, in a balanced mixture of the two lowest hyperfine states, which are loaded into a box potential $U_0$, comprising six sheets of blue-detuned light, created by two digital micromirror devices (DMDs). This produces a rectangular density profile with dimensions $(129\times84\times58)\,\mu$m, which slowly varies due to the curvature of the bias magnetic field,  Fig.~\ref{fig:boxdensity}. The average total central density is $n_0\simeq 2.6\times10^{11}$ atoms/cm$^3$, for which the Fermi energy $\epsilon_{\!F0}\equiv k_B T_F =k_B\times 0.16\,\mu$K and the Fermi speed $v_F=2.1$ cm/s. As suggested by Zhang and Yu~\cite{ZhangEnergyAbsSpectr}, we probe the linear response $\delta n(z,t)$ by employing one of the DMDs to generate a small spatially periodic optical potential that moves through the cloud at speed $v$,
\begin{equation}
\delta U(z,t)=\delta U_0[1-\epsilon\cos(qz-qvt)]\,H(vt-z),
\label{eq:drive}
\end{equation}
where $q=2\pi/\lambda$. The Heaviside envelope function $H(vt-z)$ vanishes inside the box at $t=0$. Positive light intensity requires $1-\epsilon\cos(qz-qvt)\geq 0$, so that $\epsilon\leq 1$.  For each speed $v$, $\delta U(z,t)$ is turned on for a fixed number of periods, after which an absorption image is recorded to obtain the column density. For the longest wavelength employed in the experiments, $\lambda = 30\,\mu$m, the image is taken after three periods (leading edge at $90\,\mu$m), while for the shortest wavelength $\lambda = 19\,\mu$m, imaging occurs after four periods (leading edge at $76\,\mu$m). Instead of measuring the energy input, as proposed in ref.~\cite{ZhangEnergyAbsSpectr}, we directly obtain the response $\delta n(z,t)/n_0$ from the integrated column density, which is measured five times for each $\lambda$ at several different frequencies $f\equiv v/\lambda$ from $200$ to $800$ Hz.

\begin{figure}[h!]
\centering
\includegraphics[width=2.25in]{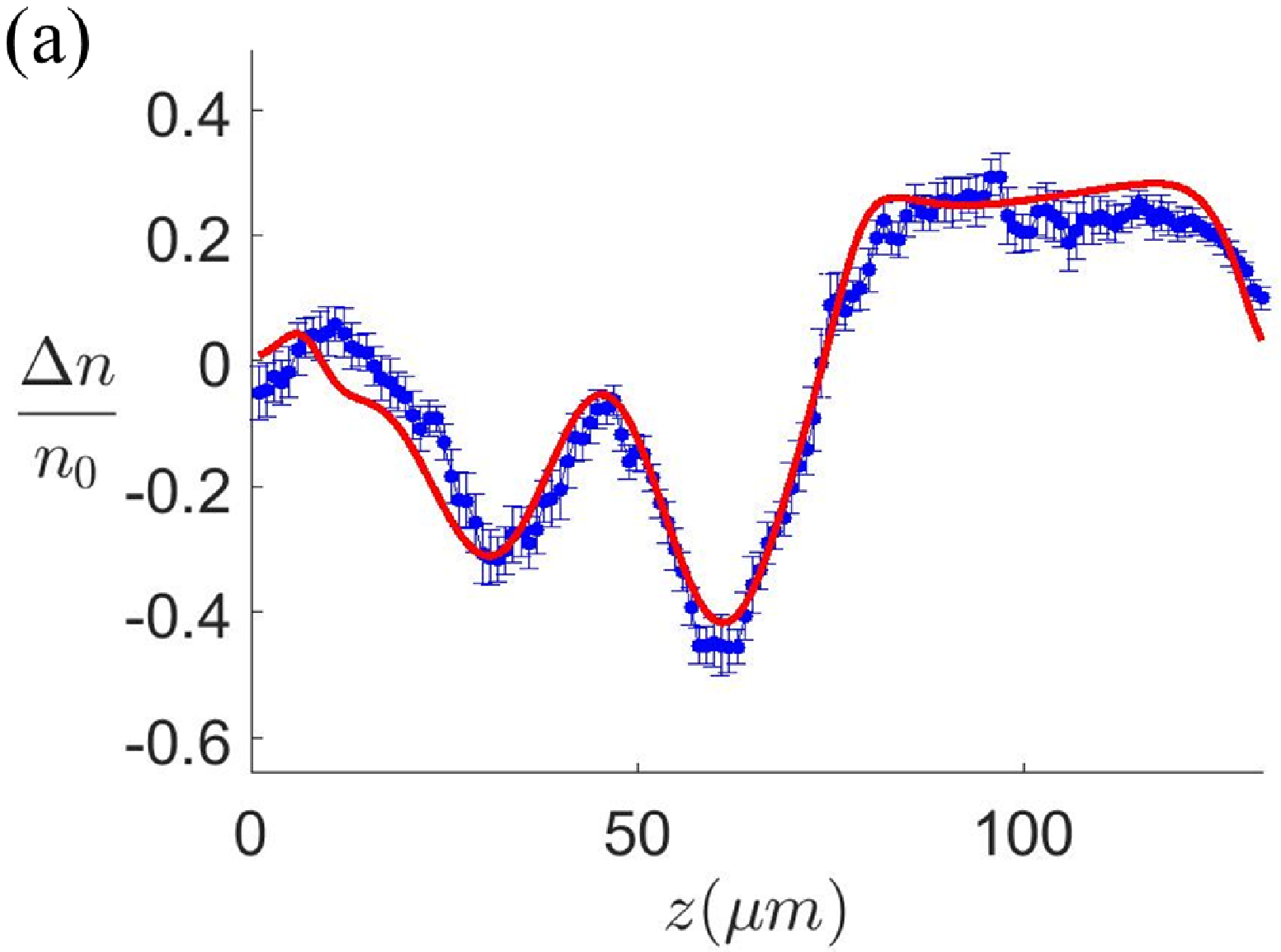}
\includegraphics[width=2.25in]{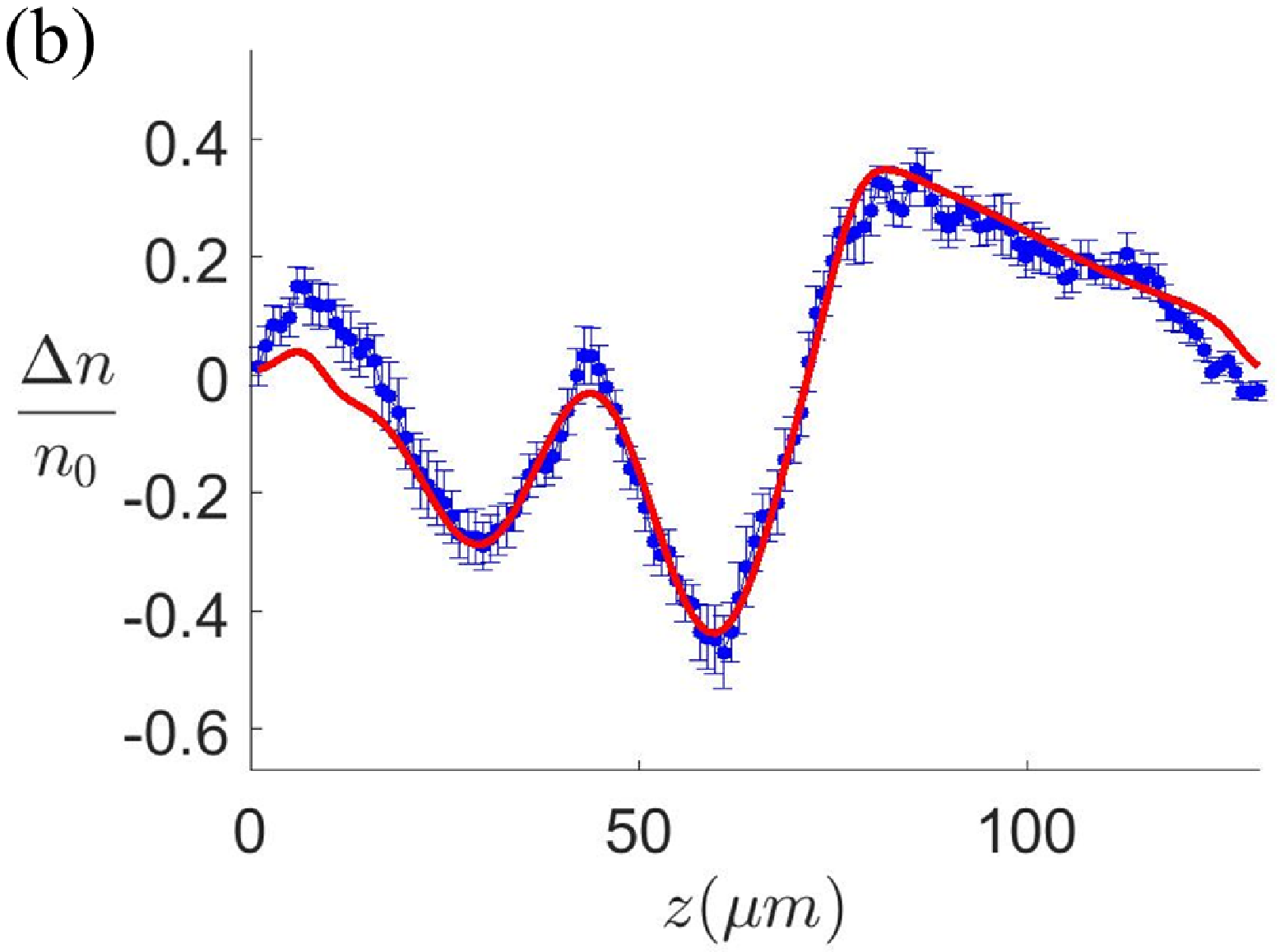}
\includegraphics[width=2.25in]{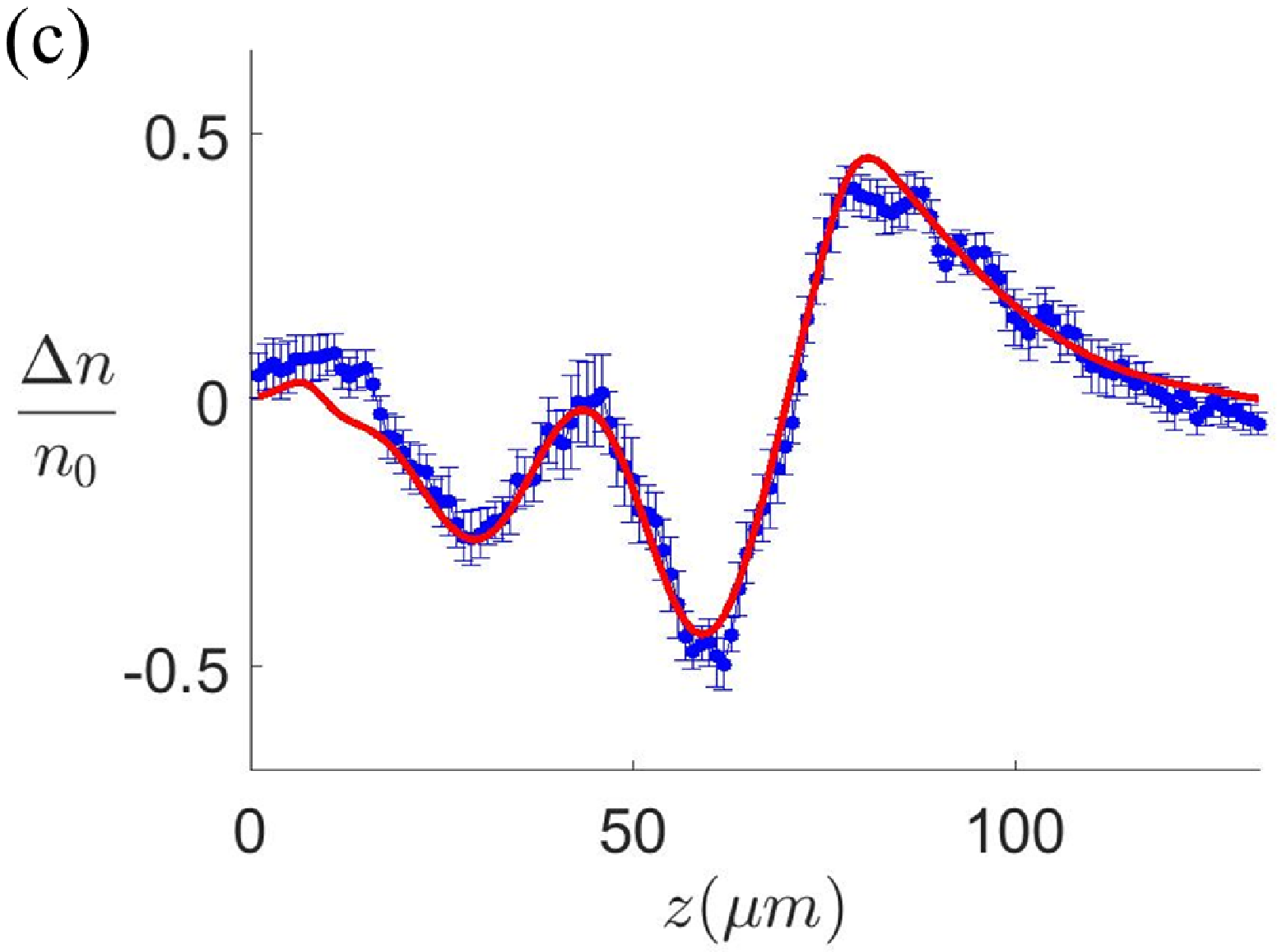}
\includegraphics[width=2.25in]{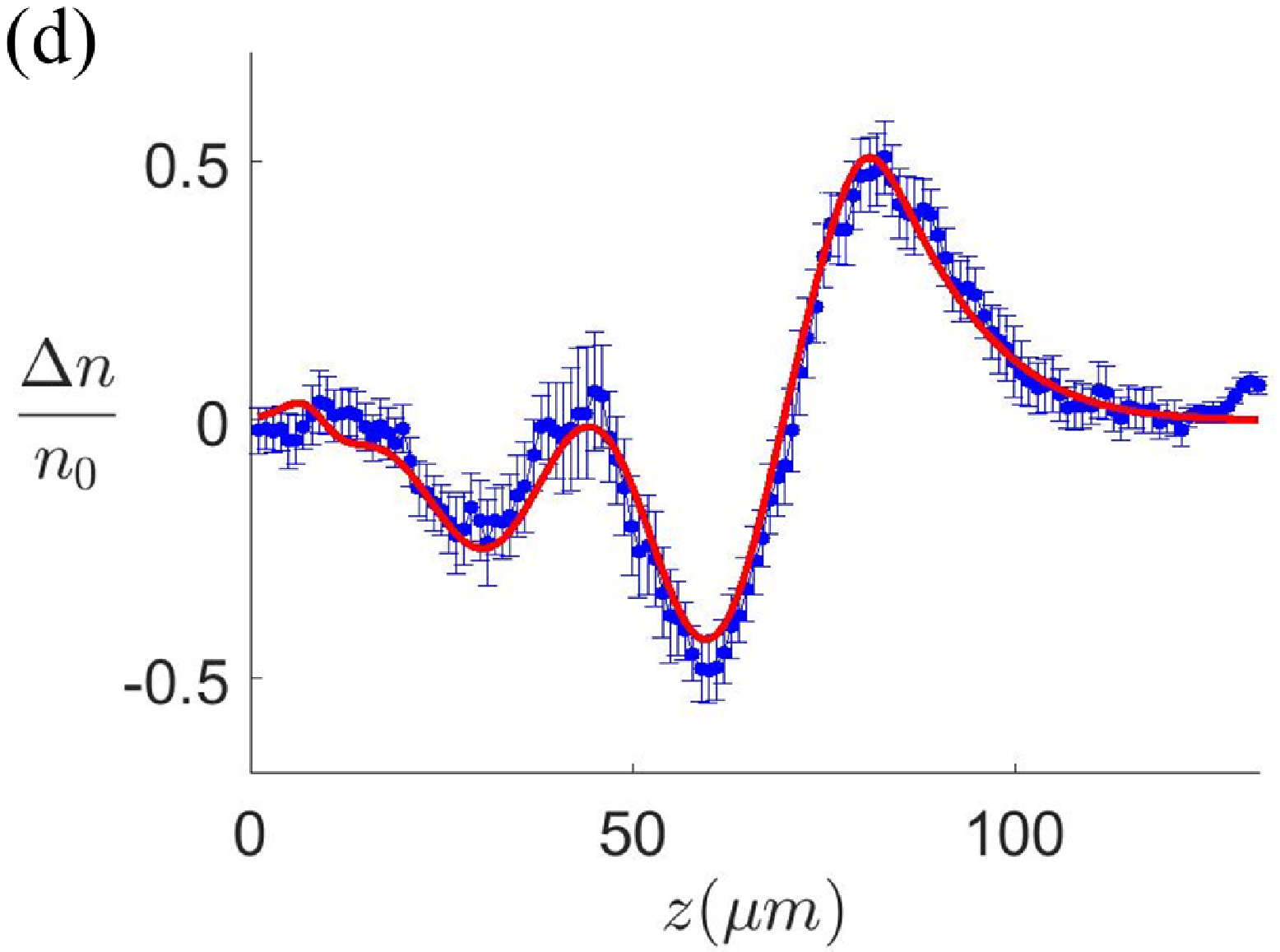}
\caption{Response to subsonic perturbations. Density change, $\delta n/n_0$, for a sinusoidal spatial perturbation with $\lambda = 30\,\mu$m, moving into the sample at a speed $v=\lambda f < c_0$ for 3 periods $1/f$. Data (blue dots); Hydrodynamic model with $c_0 = 1.3\,$cm/s, $\delta U_0 = 0.26\,\epsilon_{\!F0}$ and $\epsilon = 0.29$  (red curves) for frequencies (a) $f=200$ Hz, $v/c_0 = 0.46$ (b) $f=250$ Hz, $v/c_0 = 0.58$ (c) $f=300$ Hz, $v/c_0 = 0.69$ (d)$f=350$ Hz, $v/c_0 = 0.81$.
\label{fig:DataFits30}}
\end{figure}

\begin{figure}[h!]
\includegraphics[width=2.25in]{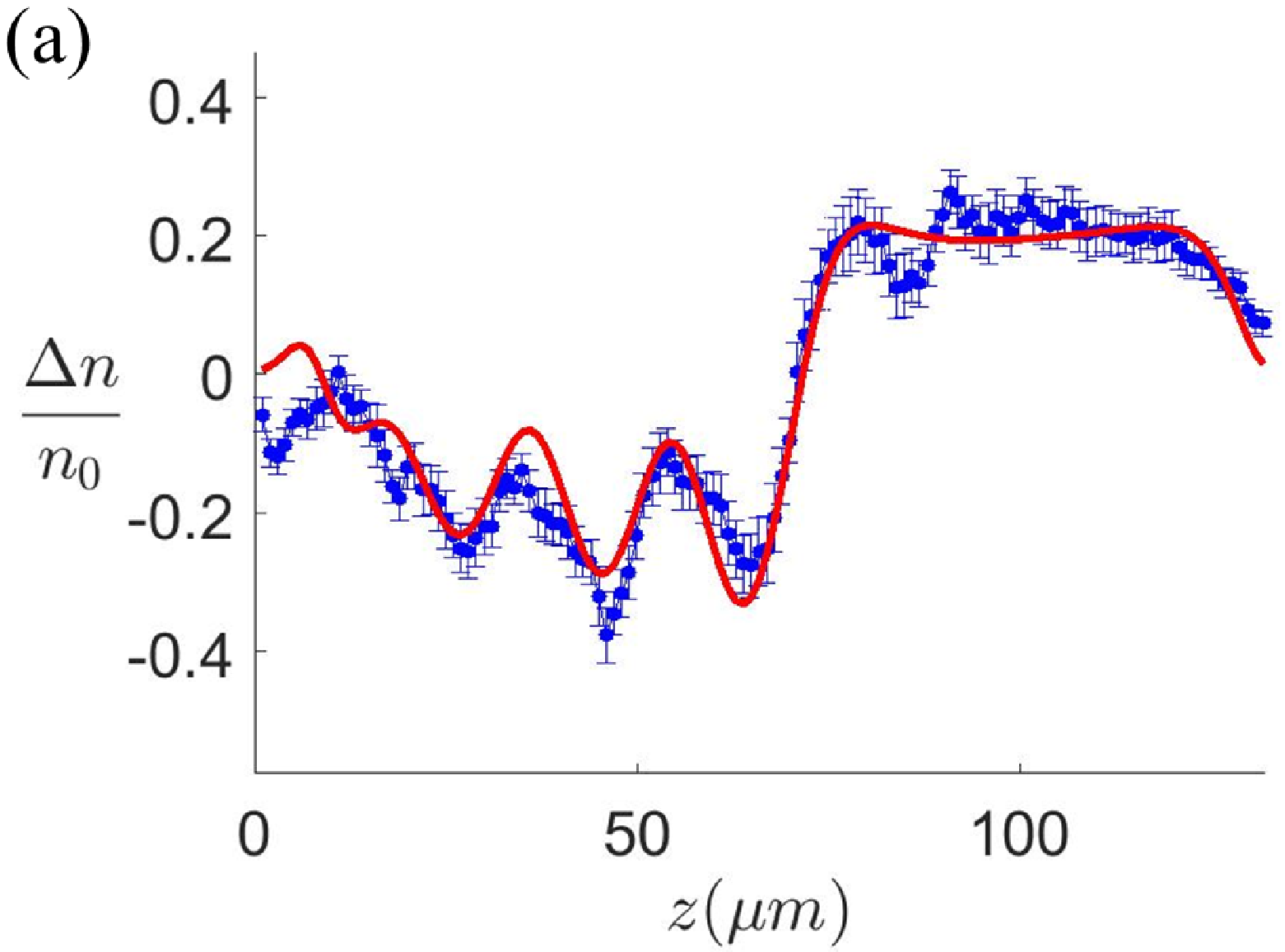}\\
\includegraphics[width=2.25in]{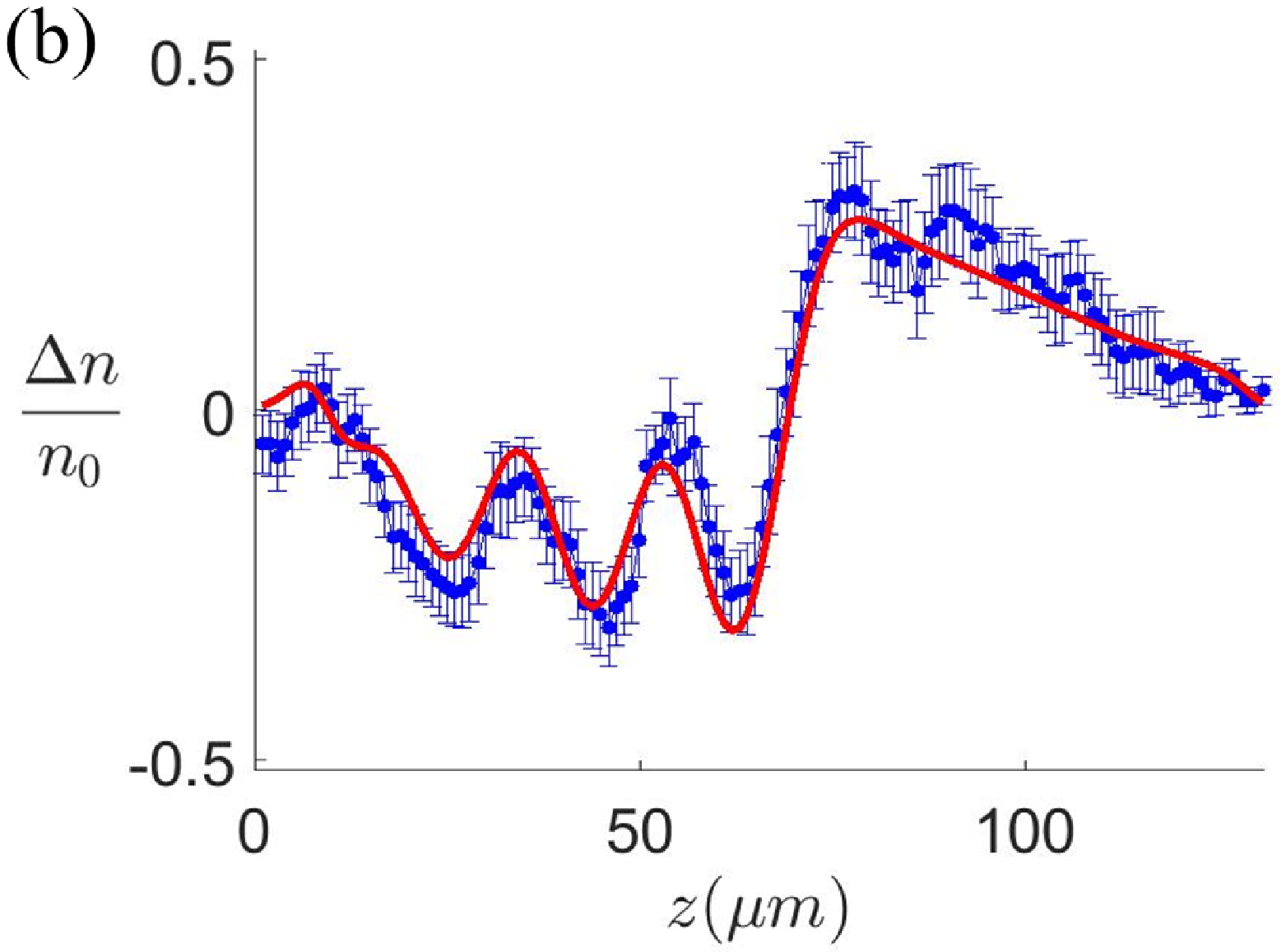}\\
\includegraphics[width=2.25in]{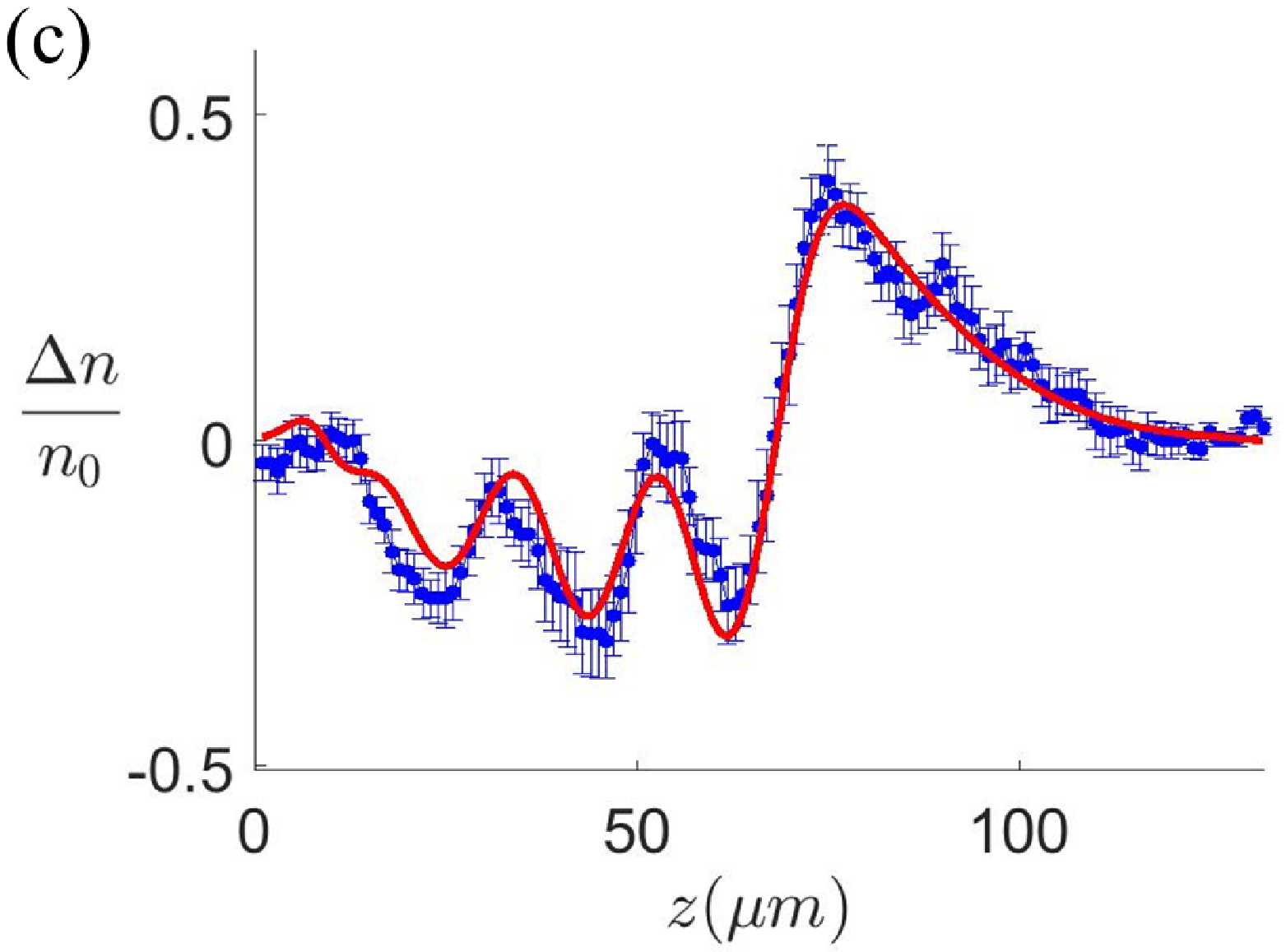}\\
\includegraphics[width=2.25in]{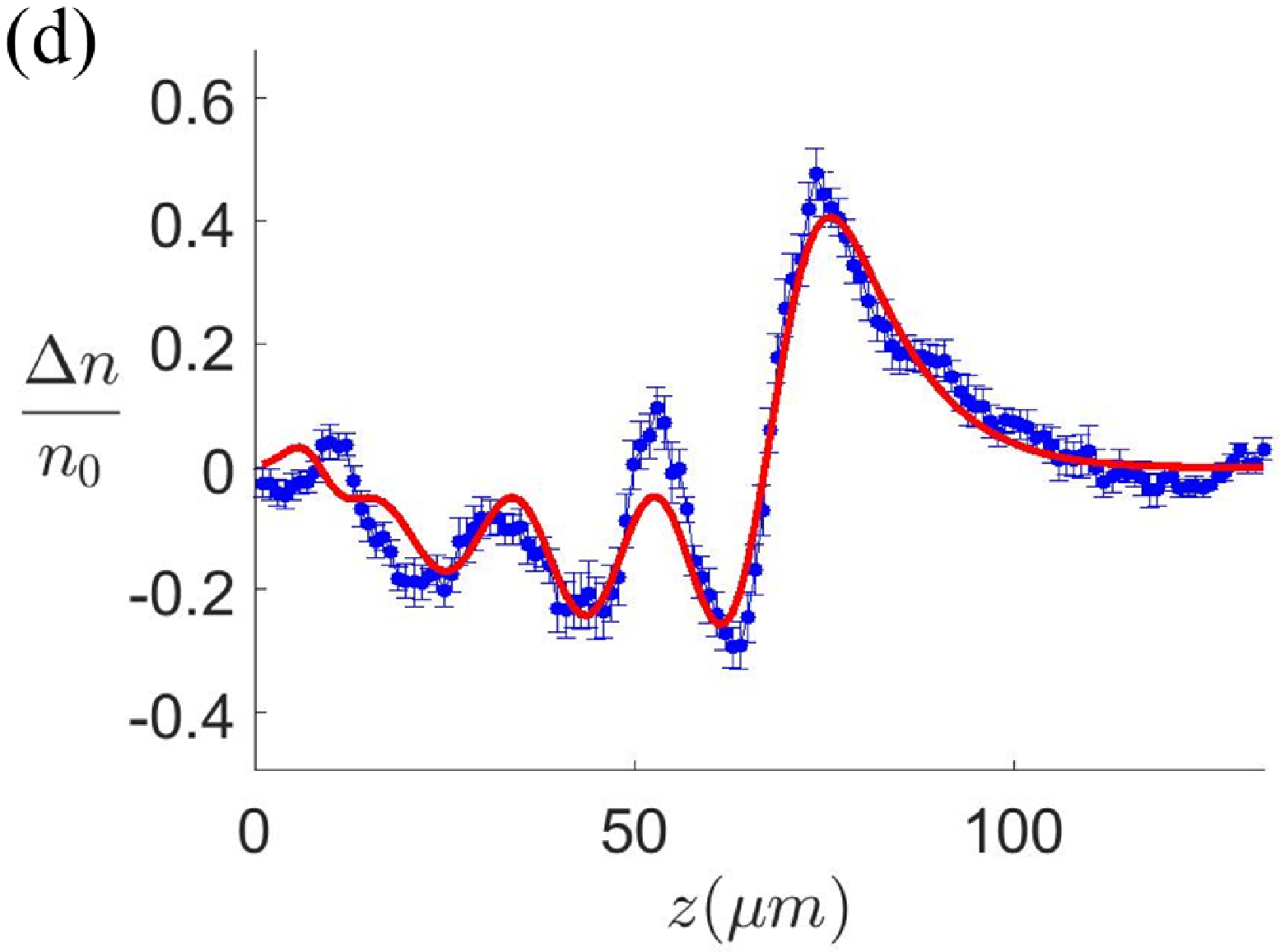}\\
\caption{Response to subsonic perturbations. Density change, $\delta n/n_0$, for a sinusoidal spatial perturbation with $\lambda = 19\,\mu$m, moving into the sample at a subsonic speed $v=\lambda f < c_0$ for 4 periods $1/f$. Data (blue dots); Hydrodynamic model with $c_0 = 1.3\,$cm/s, $\delta U_0 = 0.22\,\epsilon_{\!F0}$ and $\epsilon = 0.23$  (red curves) for frequencies (a) $f=300$ Hz, $v/c_0 = 0.44$ (b) $f=400$ Hz, $v/c_0 = 0.58$ (c) $f=500$ Hz, $v/c_0 = 0.73$ (d)$f=600$ Hz, $v/c_0 = 0.88$.
\label{fig:DataFits19}}
\end{figure}

\begin{figure}[htb]
	\includegraphics[width=3.0in]{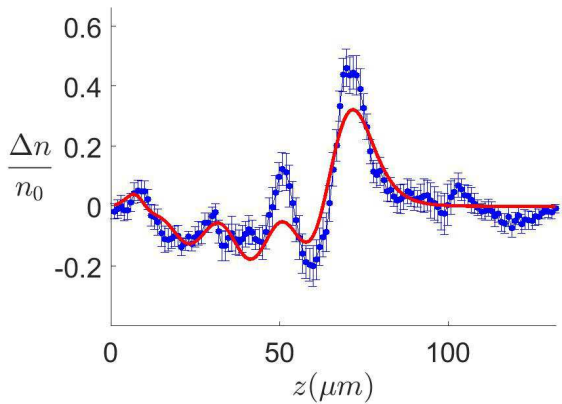}
	\caption{Response to a supersonic perturbation. Density change, $\delta n/n_0$, for a sinusoidal spatial perturbation with $\lambda = 19\,\mu$m, moving into the sample at a supersonic speed $v=\lambda f=1.17\,c_0$ for 4 periods $1/f$ and $f=800$ Hz; Data (blue dots); Hydrodynamic model with $c_0 = 1.3\,$cm/s, $\delta U_0 = 0.22\,\epsilon_{\!F0}$ and $\epsilon = 0.23$ (red curve). The thermal conductivity $\kappa_T$ cannot be extracted from the fit of the model to the supersonic data.
		\label{fig:SupersonicFit}}
\end{figure}

Figs.~\ref{fig:DataFits30},~\ref{fig:DataFits19}, and Fig.~\ref{fig:SupersonicFit} show the density response $\delta n(z,t)$ as the drive speed $v$ is varied from subsonic $v<c_0$ to supersonic $v>c_0$, where $c_0$ is the adiabatic sound speed. At low drive speeds, the leading edge of the response is nearly flat, as sound waves propagate well past the front of the driving potential. As $v$ approaches $c_0$, the amplitude of the density response increases and the leading peak narrows.

To understand the density profiles arising from the perturbation $\delta U$, we construct the coupled equations for the change in the density $\delta n$ and for the change in the entropy  per particle $\delta s_1$. The analysis is simplified for experiments in the linear response regime, where~\cite{SupportOnline}
\begin{eqnarray}
\partial_t^2\delta n-c_0^2\,(\partial_z^2\delta n+\partial_z^2\delta\tilde{s}_1)
-\frac{4}{3}\,\frac{\eta}{n_0 m}\,\partial_z^2\partial_t\delta n&=&\nonumber\\
\frac{1}{m}\,\partial_z[n_0(z)\,\partial_z\delta U+\delta n\,\partial_z U_0(z)],
\label{eq:2.5}
\end{eqnarray}
with  $m$ the atom mass. Here, $\delta\tilde{s}_1=n_0\beta\,T_0\,\delta s_1/c_P$, with $\beta$ the thermal expansivity and $T_0$ the initial sample temperature~\cite{SupportOnline}. We find
\begin{equation}
\partial_t\delta\tilde{s}_1-\frac{\kappa_T}{n_0\,c_V}\,\partial_z^2\delta\tilde{s}_1=
\frac{\kappa_T}{n_0\,c_V}\frac{c_P-c_V}{c_P}\,\partial_z^2\delta n,
\label{eq:2.6}
\end{equation}
where $c_V$ and $c_P$ are the heat capacities per particle at constant volume and at constant pressure, determined from the measured equation of state~\cite{KuThermo,SupportOnline}. On the left side of eq.~\ref{eq:2.5}, the $c_0^2$ terms arise from the pressure change $\delta p$~\cite{SupportOnline}. The $\eta$ term produces a viscous damping rate $\gamma_\eta=4\eta\, q^2/(3n_0 m)$ for the response of the density to the spatially periodic part of $\delta U(z,t)$, eq.~\ref{eq:drive}. On the right hand side of eq.~\ref{eq:2.5}, the first term  arises from the perturbing potential, with $n_0(z)$ the  background density, which varies slowly due to the bias magnetic field curvature. Here, we retain the full spatial variation of the force per unit volume~\cite{SupportOnline}, which vanishes at the box edges. In the second term, $\partial_zU_0(z)$ is the force from the box potential. We determine $\partial_z U_0(z)$ from $n_0(z)$, which is measured in equilibrium~\cite{SupportOnline}.

In addition to the shear viscosity, $\delta n(z,t)$ carries information about the thermal conductivity $\kappa_T$, which sets the relaxation rate, $\gamma_\kappa=\kappa_T q^2/(n_0c_V)$  in eq.~\ref{eq:2.6}, of the spatially periodic temperature profile that is imprinted by $\delta U(z,t)$,  eq.~\ref{eq:drive}. For a high speed $v$, the wave frequency $q v>>\gamma_\kappa$. Then $\partial_t\delta\tilde{s}_1$ dominates in eq.~\ref{eq:2.6} and $\delta\tilde{s}_1\simeq 0$, yielding $(\partial_t^2-c_0^2\,\partial_z^2)\,\delta n$ on the left side of eq.~\ref{eq:2.5}. In this case, the compression is adiabatic, and sound waves propagate at the speed $c_0$. In the opposite limit of a low speed $v$, the wave frequency $qv<<\gamma_\kappa$. Eq.~\ref{eq:2.6} shows that $\partial_z^2\delta\tilde{s}_1\simeq-(c_P-c_V)/c_P\,\partial_z^2 \delta n$, yielding  $(\partial_t^2-c_T^2\,\partial_z^2)\,\delta n$  in eq.~\ref{eq:2.5}, with $c_T=c_0\sqrt{c_V/c_P}$. Then, the compression is isothermal and sound waves propagate at the isothermal sound speed $c_T$~\cite{SupportOnline}.

To model the normal fluid data,  $c_0$ is used as a fit parameter. The fitted $c_0$ then serves as a thermometer, as the reduced temperature $\theta_0 = T_0/T_F$ of the gas is monotonically related to $c_0/v_F$ in the normal fluid regime~\cite{SupportOnline}. With $\theta_0$ determined, $c_V$ and $c_P$ are then fixed by the measured equation of state~\cite{KuThermo,SupportOnline}. Further, $\theta_0$ determines the shear viscosity $\eta$ as discussed below.

Our analysis benefits from recent progress in determining the local shear viscosity of the normal fluid from hydrodynamic expansion experiments~\cite{JosephShearNearSF,BluhmSchaeferLocalViscosity}. Extraction of $\eta$ is simplified in expansion measurements, because the temperature gradient is negligible~\cite{TempGrad} so that the thermal conductivity $\kappa_T$ can be neglected.  The most complete data for the shear viscosity have been obtained from the aspect ratio of expanding cigar-shaped clouds, measured at a fixed time $t$ after release from an optical trap as a function of the cloud energy~\cite{JosephShearNearSF}. The latest hydrodynamic analysis utilizes an anisotropic pressure model, which properly interpolates between the hydrodynamic behavior in the dense regions of the cloud and the free streaming ballistic expansion near the cloud edges~\cite{BluhmSchaeferLocalViscosity}.  The new analysis yields an expansion of the local shear viscosity in powers of the diluteness $n\lambda_T^3$,
\begin{equation}
\eta=\eta_0\frac{(mk_BT)^{3/2}}{\hbar^2}\,[1+\eta_2(n\lambda_T^3)+...],
\label{eq:1.1}
\end{equation}
where $\lambda_T=h/\sqrt{2\pi mk_BT}$ is the thermal wavelength and $n$ is the total density for a balanced two-component mixture. Fits to the expansion data yield $\eta_0=0.265(20)$, in excellent agreement with the variational result obtained from the two-body Boltzmann equation for a unitary gas, eq.~\ref{eq:etaHighT}, $\eta_0=15/(32\sqrt{\pi})=0.26446$~\cite{BluhmSchaeferLocalViscosity}. This confirms that the data and the analysis properly reproduce the high temperature limit, which is independent of the density. The next order term is independent of the temperature, with $\eta_2=0.060(20)$, while the $\eta_3(n\lambda_T^3)^2$ term is  negligible. Remarkably, the first two terms fit the expansion data down to temperatures just above the superfluid transition. We therefore use eq.~\ref{eq:1.1} as in input for eq.~\ref{eq:2.5}, where $\eta/(n_0 m)\equiv\alpha(\theta_0)\,\hbar/m$ and $\alpha(\theta_0)=\alpha_0\,\theta_0^{3/2}+\alpha_2$, with $\alpha_0=(3\pi^2/\sqrt{8})\,\eta_0=2.77$ and $\alpha_2=(2\pi)^{3/2}\eta_0 \eta_2=0.25$.

The data  are modeled by numerically integrating eqs.~\ref{eq:2.5}~and~\ref{eq:2.6} using four fit parameters, $c_0$, $\delta{U_0}$, and $\epsilon$, given in the figure captions, and $\kappa_T$, which is discussed below. These parameters are extracted by minimizing $\chi^2$ in the central region of the data away from the less dense edges.  The fits are done one parameter at a time across all frequencies for a global best fit, with  $c_P$, $c_V$, and $\eta$ determined by $\theta_0(c_0)$. This process is repeated until variation in the parameters no longer results in improvement.  The sensitivity to $\epsilon$ is greatest where the  density response shows periodic modulation, while $c_0$ is dominant in the shape of the leading edges,  Figs.~\ref{fig:DataFits30}~and~\ref{fig:DataFits19}. The fitted $\delta{U_0}$ values are consistent with the value $0.2\,\epsilon_{\!F0}$ estimated from the expected modulation depth and the maximum box potential $4.5\,\epsilon_{\!F0}$~\cite{SupportOnline}. Blurring arising from the imaging resolution $\simeq 3.5\,\mu$m, causes the fitted $\epsilon$ for the $19\,\mu$m data to be smaller than for the $30\,\mu$m data. We find that the model captures both the amplitudes and shapes of the density response $\delta n(z,t)/n_0$ for all of the frequencies, Figs.~\ref{fig:DataFits30}~and~\ref{fig:DataFits19}.

From the $\chi^2$ fits for both $\lambda = 19\,\mu$m and for $\lambda=30\,\mu$m, we obtain $c_0=1.30$ cm/s, consistent with sound speed measurements in the uniform cloud, which gives $1.40$ cm/s. The measured $c_0$ determines $\theta_0=0.50$~\cite{SupportOnline}.  The temperature was not further increased, because the box potential was not strong enough to confine the gas at significantly higher temperature.

We see that the quality of fits decreases as the speed approaches the adiabatic sound speed, $v/c_0=0.88$, Fig.~\ref{fig:DataFits19} (d).  In the supersonic regime, Fig.~\ref{fig:SupersonicFit}, we find that the fit of the linear hydrodynamic model to the density response is poor, and the thermal conductivity cannot be reliably extracted from the model for any perturbation moving faster than the adiabatic sound speed. We estimate that the hydrodynamic relaxation time is $\tau=0.13$ ms~\cite{SupportOnline}, which is fast compared to the period of 1.25 ms at the frequency $f=800$ Hz used to observe the supersonic response. However, in the supersonic regime, it is possible that the increasing density gradients produce weak shock waves, which are not included in our model.

Sensitivity to $\kappa_T$ is enabled by measurement at subsonic speeds, as
the frequency $v/\lambda$ can be less than the relaxation rate $\gamma_\kappa=\kappa_T q^2/(n_0c_V)$. Using eq.~\ref{eq:kappaHighT}, with $\theta_0=0.50$, we find $\gamma_\kappa=2\pi\times 760$ Hz for $\lambda=19\,\mu$m and $\gamma_\kappa=2\pi\times 305$ Hz at $\lambda=30\,\mu$m. The fits to the trailing edge of the leading peak rise more sharply for larger $\kappa_T$, because the density response propagates closer to the isothermal sound speed $c_T<c_0$ for large $\gamma_\kappa$ and lags behind the leading peak to cause a larger disturbance. From the fits to the subsonic data, we find $\kappa_T=1.14(17)\times(15/4)(k_B/m)\hbar\,n_0$ at $\theta_0=0.50$.

The fitted thermal conductivity at $\theta_0=0.50$ for the unitary Fermi gas can be compared with the variational calculations~\cite{BrabySchaeferThermalCond}. As noted above, the high temperature limit of the shear viscosity fits the expansion data down to temperatures just above the superfluid transition. For this reason, we compare the fit value of $\kappa_T/\eta$ to the predicted high temperature ratio,  eq.~\ref{eq:kappaHighT}, $\kappa_T/\eta=(15/4)(k_B/m)$. This ratio holds for the unitary gas and for an energy-independent s-wave scattering cross section~\cite{BrabySchaeferThermalCond}, and is identical to the predictions and measurements for rare gases in the Boltzmann limit~\cite{ReifViscosity,KittelViscosity}.  With the viscosity from the expansion data, as used in the fits, $\eta = 1.23\,\hbar\,n_0$ at $\theta_0=0.5$, we find $\kappa_T/\eta=0.93(14)\,\times(15/4)(k_B/m)$, close to the ratio predicted in the high temperature limit. Finally, we determine the Prandtl number,  $Pr=(c_P/m)\,\eta/\kappa_T$~\cite{BrabySchaeferThermalCond,HuTwoFluidPRA}. For $\theta_0=0.5$, we find $c_P =1.68\,k_B$ from the equation of state~\cite{KuThermo,SupportOnline}, yielding $Pr = 0.48(8)$, which can be compared to the high temperature limit $Pr=2/3$, obtained from eq.~\ref{eq:kappaHighT} with $c_P =5/2\,k_B$.

In conclusion, we have directly measured the hydrodynamic response of a unitary Fermi gas subject to a moving spatially periodic perturbation. The measured density perturbations validate a linear response model that incorporates the measured box potential, enabling predictions beyond the approximation of an infinite medium. From the low frequency response, we obtain an estimate of the thermal conductivity of the normal fluid that is consistent with recent predictions. Future measurements in improved box potentials will permit studies of the thermal conductivity at higher temperatures, enabling more precise comparison with benchmark variational calculations. Further, this new method will enable measurement of the thermal conductivity and shear viscosity for imbalanced mixtures in nearly uniform gases, where the transport properties are predicted to change~\cite{CaiViscUniformFGPopImbal}.

{\it Note added:} Shortly after submission of the present paper, a related study appeared on arXiv, measuring universal sound diffusion in a strongly interacting Fermi gas~\cite{MZSoundDiffusion}.

Primary support for this research is provided by the Physics Divisions of the National Science Foundation (PHY-1705364) and the Air Force Office of Scientific Research (FA9550-16-1-0378). Additional support for the JETlab atom cooling group has been provided by the Physics Division of the Army Research Office (W911NF-14-1-0628) and by the Division of Materials Science and Engineering, the Office of Basic Energy Sciences, Office of Science, U.S. Department of Energy (DE-SC0008646).

$^*$Corresponding author: jethoma7@ncsu.edu

%

\widetext
\appendix
\section{Supplemental Material}
\label{sec:supplement}
In this supplemental material, we derive the hydrodynamic linear response model, including the thermodynamics and the method of determining the forces arising from the box potential.
\subsection{Hydrodynamic linear response for a normal fluid.}
Our analysis is simplified for a normal  Fermi gas, which is a single component fluid with a mass density is $\rho\equiv n\,m$, where $n$ is the total particle density (we assume a 50-50 mixture of two components) and $m$ is the atom mass. $\rho({\mathbf{r}},t)$ satisfies the continuity equation,
\begin{equation}
\partial_t\rho +\partial_i(\rho\,v_i)=0,
\label{eq:1.3}
\end{equation}
where a sum over $i=x,y,z$ is implied. The mass flux (momentum density) is $\rho\,v_i$, with $v_i({\mathbf{r}},t)$ the velocity field.

The momentum density and corresponding momentum flux $\rho\,v_i v_j$ obey
\begin{equation}
\partial_t (\rho\,v_i) +\partial_j (\rho\,v_i v_j)=-\partial_i p -n\,\partial_i U+\partial_j(\eta\,\sigma_{ij}),
\label{eq:1.5}
\end{equation}
Here, $-\partial_i p-n\,\partial_i U$ is the force per unit volume arising from the pressure $p$ and the externally applied potential $U({\mathbf{r}},t)$. The last term describes the dissipative forces, which arise only from the shear viscosity  $\eta$, as the bulk viscosity vanishes for a unitary gas. The shear stress tensor is $\sigma_{ij}\equiv\partial_iv_j+\partial_jv_i-2\delta_{ij}\nabla\cdot{\mathbf{v}}/3$. Taking the divergence of eq.~\ref{eq:1.5}, and using eq.~\ref{eq:1.3}, we immediately obtain
\begin{equation}
-\partial_t^2\rho+\partial_i\partial_j (\rho\,v_i v_j)=-\partial_i^2 p-\partial_i(n\,\partial_iU)+\partial_i\partial_j(\eta\,\sigma_{ij}).
\label{eq:2.7}
\end{equation}

We are interested in the linear hydrodynamic response to a perturbing external potential $\delta U({\mathbf{r}},t)$, which leads to first order changes in the density $\delta n({\mathbf{r}},t)$ and pressure $\delta p({\mathbf{r}},t)$,
\begin{eqnarray}
n({\mathbf{r}},t)&=&n_0({\mathbf{r}})+\delta n({\mathbf{r}},t)\nonumber\\
p({\mathbf{r}},t)&=&p_0({\mathbf{r}})+\delta p({\mathbf{r}},t)\nonumber\\
U({\mathbf{r}},t)&=&U_0({\mathbf{r}})+\delta U({\mathbf{r}},t).
\label{eq:3.1}
\end{eqnarray}
Here, $n_0({\mathbf{r}})$ and $p_0({\mathbf{r}})$ are the equilibrium (time independent) density and pressure arising from confinement in the box trap potential, $U_0({\mathbf{r}})$. In equilibrium, the velocity field ${\mathbf v}_0({\mathbf{r}},t)=0$ and eq.~\ref{eq:1.5} requires balance of the forces per unit volume arising from the box trap and the pressure,
\begin{equation}
-\nabla p_0({\mathbf{r}})-n_0({\mathbf{r}})\nabla U_0({\mathbf{r}})=0.
\label{eq:1.3b}
\end{equation}
Substituting eq.~\ref{eq:3.1} into  eq.~\ref{eq:2.7} and retaining terms to first order in small quantities, we obtain
\begin{equation}
\partial_t^2\delta n=\frac{1}{m}\nabla^2\,\delta p+\frac{1}{m}\nabla\cdot[n_0({\mathbf{r}})\,\nabla\delta U+\delta n\,\nabla U_0]+\frac{4}{3}\frac{\eta}{n_0 m}\nabla^2\partial_t\delta n.
\label{eq:3.7}
\end{equation}
Here, the second term on the left side of eq.~\ref{eq:2.7} is negligible, as the velocity field is first order in small quantities. To obtain the last term in eq.~\ref{eq:3.7}, we have used the fact that the dissipative force is small compared to the conservative forces and that the density $n_0$ slowly varies in the region of interest. Then we can ignore the spatial derivatives of $\eta$ and $n_0$, taking $\partial_i\partial_j(\eta\,\sigma_{ij})\simeq\eta\,\partial_i\partial_j\sigma_{ij}=4\eta/3\,\nabla^2(\nabla\cdot{\mathbf{v}})$. The velocity field is eliminated using $\nabla\cdot{\mathbf{v}}\simeq -\partial_t\delta n/n_0$, which follows from  eq.~\ref{eq:1.3}.

We retain the spatial dependence of the driving force per unit volume $n_0({\mathbf{r}})\,\nabla\delta U$ in eq.~\ref{eq:3.7}, to assure that it vanishes smoothly at the cloud edges. Here, the perturbing potential $\delta U$ is controlled in the experiments. The force arising from the box potential $-\nabla U_0$ is determined from eq.~\ref{eq:1.3b}, using the equation of state $p_0[n_0({\mathbf{r}}),T]$, where the temperature $T$ is spatially constant in equilibrium. Then,
\begin{equation}
\nabla U_0({\mathbf{r}})=-\left(\frac{\partial p_0}{\partial n_0}\right)_{\!\!T}\frac{\nabla n_0({\mathbf{r}})}{n_0({\mathbf{r}})}.
\label{eq:3.6}
\end{equation}
Eq.~\ref{eq:3.6}, with $(\partial p/\partial n)_T =(c_V/c_P)\,m\,c_0^2$,  yields $\nabla U_0({\mathbf{r}})$, from the equilibrium density profile, where $c_V$ and $c_P$ are the heat capacities per particle at constant volume and pressure, and $c_0$ is the adiabatic sound speed (see Fig.~\ref{fig:soundspeed}).

To proceed further, we need an expression for the first order pressure change, $\delta p$, which is determined from eq.~\ref{eq:6.3T} by setting $n=n_0$ and $T=T_0$,
\begin{equation}
\delta p=m\,c_0^2\,\delta n+mc_0^2\beta\,\frac{T_0}{c_P}\,n_0\,\delta s_1.
\label{eq:6.3}
\end{equation}
As discussed in the thermodynamics section below, the first term in eq.~\ref{eq:6.3} arises from adiabatic compression and in the second term, $\delta s_1$ is the first order change in the entropy per particle, with $\beta$ is the thermal expansivity. Inserting eq.~\ref{eq:6.3} into eq.~\ref{eq:3.7}, we obtain
\begin{equation}
\partial_t^2\delta n-c_0^2\nabla^2\,\delta n-c_0^2\,\beta n_0\frac{T_0}{c_P}\nabla^2\delta s_1-\frac{4}{3}\frac{\eta}{n_0 m}\nabla^2\partial_t\delta n=\frac{1}{m}\nabla\cdot[n_0({\mathbf{r}})\,\nabla\delta U+\delta n\,\nabla U_0].
\label{eq:4.6}
\end{equation}

The evolution equation for $\delta s_1$ is determined by the heating rate per unit volume, $\delta\dot{q}$,
\begin{equation}
T_0(\partial_t+{\mathbf v}\cdot\nabla)\,\delta s_1=\frac{\delta\dot{q}}{n_0},
\label{eq:4.2}
\end{equation}
where $T_0$ is the initial, spatially uniform, temperature.
The heating rate arising from the shear viscosity is second order in $v_i$, which is negligible compared to the heating rate arising from heat conduction. Hence, $\delta\dot{q}\simeq -\nabla\cdot(-\kappa_T\nabla\delta T)\simeq\kappa_T\nabla^2\delta T$, where we neglect the spatial derivatives of $\kappa_T$. Here, $\delta T$ is the first order temperature change, which is determined from eq.~\ref{eq:6.5T} by setting $n=n_0$ and $T=T_0$,
\begin{equation}
\delta T=\frac{T_0}{c_V}\delta s_1 + mc_0^2\,\beta\frac{T_0}{c_P}\frac{\delta n}{n_0}.
\label{eq:4.4}
\end{equation}
Retaining only first order terms in eq.~\ref{eq:4.2} and using eq.~\ref{eq:4.4}, we have
\begin{equation}
\partial_t\delta s_1-\frac{\kappa_T}{n_0 c_V}\nabla^2\delta s_1=\frac{\kappa_Tmc_0^2}{n_0^2c_P}\beta\,\nabla^2\delta n.
\label{eq:4.5}
\end{equation}

We define $\delta \tilde{s}_{1}\equiv\beta T_0\,n_0\delta s_1/c_P$, which has a dimension of density, and take $\nabla^2\rightarrow\partial_z^2$ in eqs.~\ref{eq:4.6}~and~\ref{eq:4.5} to obtain
\begin{equation}
\partial_t^2\delta n-c_0^2(\partial_z^2\,\delta n+\partial_z^2\delta \tilde{s}_{1})-\frac{4}{3}\frac{\eta}{n_0 m}\partial_z^2\partial_t\delta n=\frac{1}{m}\,\partial_z[n_0(z)\,\partial_z\delta U+\delta n\,\partial_z U_0].
\label{eq:1.4}
\end{equation}
and
\begin{equation}
\partial_t\delta\tilde{s}_{1}-\frac{\kappa_T}{n_0 c_V}\,\partial_z^2\delta\tilde{s}_{1}=\frac{\kappa_T}{n_0c_V}\frac{c_P-c_V}{c_P}\,\partial_z^2\delta n,
\label{eq:1.5b}
\end{equation}
where we have used $mc_0^2\beta^2T_0/c_P=(c_P-c_V)/c_V$, as shown in below, eq.~\ref{eq:5.7T}.

For the numerical integration, it is convenient to define the dimensionless variables, $\delta\tilde{n}\equiv\delta n/n_0$ and $\tilde{n}_0(z)\equiv n_0(z)/n_0$, where $n_0$ is the average total density. Similarly $\delta\tilde{s}_{10}=\delta\tilde{s}_1/n_0=\beta T_0\,\delta s_1/c_P$. Further, we define  $\delta\tilde{U}(z,t)\equiv\delta U(z,t)/\epsilon_{F0}$ and $\tilde{U}_0(z)\equiv U_0(z)/\epsilon_{F0}$, where the local Fermi energy, evaluated for the density $n_0$, is $\epsilon_{F0}=mv_{F0}^2/2=\hbar^2(3\pi^2n_0)^{2/3}/(2m)$. In terms of these variables, eqs.~\ref{eq:1.4}~and~\ref{eq:1.5b} become,
\begin{equation}
\partial_t^2\delta\tilde{n}-c_0^2(\partial_z^2\,\delta\tilde{n}+\partial_z^2\delta\tilde{s}_{10})-\frac{4}{3}\frac{\eta}{n_0 m}\partial_z^2\partial_t\delta\tilde{n}=\frac{v_{F0}^2}{2}\,\partial_z[\tilde{n}_0(z)\,\partial_z\delta\tilde{U}+\delta \tilde{n}\,\partial_z\tilde{U}_0]
\label{eq:1.7a}
\end{equation}
and
\begin{equation}
\partial_t\delta\tilde{s}_{10}-\frac{\kappa_T}{n_0 c_V}\,\partial_z^2\delta\tilde{s}_{10}=\frac{\kappa_T}{n_0c_V}\frac{c_P-c_V}{c_P}\,\partial_z^2\delta\tilde{n}.
\label{eq:1.8a}
\end{equation}
We show below how the box force $-\partial_z\tilde{U}_0(z)$ and the box potential $\tilde{U}_0(z)$ are determined from the measured density profile $\tilde{n}_0(z)$.

For the fits, we take the perturbation to have the form,
\begin{equation}
\delta\tilde{U}(z,t)=\delta\tilde{U}_0\,[1-\epsilon\cos(qz-qvt)]\,\frac{1 - \tanh[(z-v t)/w_z]}{2}.
\label{eq:perturb}
\end{equation}
In eq.~\ref{eq:perturb}, $\delta\tilde{U}_0$ is the strength of the perturbation in units of $\epsilon_{F0}$ and $\epsilon$ is modulation depth in the plane of the atoms. The last factor is unity for $z<vt$ and vanishes for $z>vt$, where the width $w_z$ (typically a few microns) assures a smooth transition for the numerical integration.

\subsection{Thermodynamics}
\label{sec:thermo}
\subsubsection{Thermodynamic relations}

For later use in determining the pressure change $\delta p$  and temperature change $\delta T$, we begin by deriving the well-known formula relating the heat capacity at constant pressure $c_P=T(\partial s_1/\partial T)_P$ to the heat capacity at constant volume $c_V=T(\partial s_1/\partial T)_V$. In the following, we define the volume per particle $V_1=1/n$, where $n$ is the local density, and determine the heat capacities per particle from the entropy per particle $s_1(p,T)$.
\begin{equation}
c_V=T\left(\frac{\partial s_1}{\partial T}\right)_{\!\!V_1}=T\left(\frac{\partial s_1}{\partial T}\right)_{\!\!P}+T\left(\frac{\partial s_1}{\partial p}\right)_{\!\!T}\left(\frac{\partial p}{\partial T}\right)_{\!\!V_1}
\label{eq:1.3T}
\end{equation}
Rewriting eq.~\ref{eq:1.3T} using the Maxwell relation $(\partial s_1/\partial p)_T=-(\partial V_1/\partial T)_p$, obtained from the Gibb's free energy, we have
\begin{equation}
c_V=c_P-T\left(\frac{\partial V_1}{\partial T}\right)_{\!\!P}\left(\frac{\partial p}{\partial T}\right)_{\!\!V_1}.
\label{eq:1.7T}
\end{equation}
This result can be written in terms of the thermal expansivity $\beta$,
\begin{equation}
\beta\equiv \frac{1}{V_1}\left(\frac{\partial V_1}{\partial T}\right)_{\!\!P}.
\label{eq:2.1T}
\end{equation}
Using $dp(V_1,T)=0$, we have
\begin{equation}
\left(\frac{\partial p}{\partial T}\right)_{\!\!V_1}=-\left(\frac{\partial p}{\partial V_1}\right)_{\!\!T}\left(\frac{\partial V_1}{\partial T}\right)_{\!\!P}=
\frac{1}{V_1^2}\left(\frac{\partial V_1}{\partial T}\right)_{\!\!P}\left(\frac{\partial p}{\partial n}\right)_{\!\!T}
\label{eq:2.4T}
\end{equation}
In the last step, we have used $n=1/V_1$ so that $(\partial p/\partial V_1)_T=-1/V_1^2\,(\partial p/\partial n)_T$. Inserting eq.~\ref{eq:2.4T}
into eq.~\ref{eq:1.7T} and using eq.~\ref{eq:2.1T}, we obtain the desired relation
\begin{equation}
c_V=c_P-\beta^2T\left(\frac{\partial p}{\partial n}\right)_{\!\!T}.
\label{eq:2.8T}
\end{equation}

For later use, we rewrite the  Maxwell relation $(\partial s_1/\partial p)_T=-(\partial V_1/\partial T)_p$ as
\begin{equation}
\left(\frac{\partial s_1}{\partial p}\right)_{\!T}=-\beta V_1
\label{eq:2.2T}
\end{equation}
and eq.~\ref{eq:2.4T}, using $n=1/V_1$ and $(\partial p/\partial T)_{V_1}=(\partial p/\partial T)_n$, as
\begin{equation}
\left(\frac{\partial p}{\partial T}\right)_{\!\!n}=\beta\, n \left(\frac{\partial p}{\partial n}\right)_{\!\!T}.
\label{eq:3.1T}
\end{equation}

Next, we relate the isothermal sound speed $c_T$ to the adiabatic speed of sound $c_0$,
\begin{equation}
c_0^2\equiv\frac{1}{m}\left(\frac{\partial p}{\partial n}\right)_{\!\!s_1}.
\label{eq:4.1aT}
\end{equation}
\begin{equation}
c_T^2\equiv\frac{1}{m}\left(\frac{\partial p}{\partial n}\right)_{\!\!T}.
\label{eq:4.1bT}
\end{equation}

Writing $p(n,s_1)=p[n,T(n,s_1)]$, we have
\begin{equation}
\left(\frac{\partial p}{\partial n}\right)_{\!\!s_1}=\left(\frac{\partial p}{\partial n}\right)_{\!\!T}+\left(\frac{\partial p}{\partial T}\right)_{\!\!n}\left(\frac{\partial T}{\partial n}\right)_{\!\!s_1}.
\label{eq:4.2T}
\end{equation}
We rewrite the last factor on the right hand side, using $ds_1(p,T)=0$  to obtain
\begin{eqnarray}
0&=&\left(\frac{\partial s_1}{\partial p}\right)_{\!\!T}\left(\frac{\partial p}{\partial n}\right)_{\!\!s_1}+\left(\frac{\partial s_1}{\partial T}\right)_{\!\!P}\left(\frac{\partial T}{\partial n}\right)_{\!\!s_1}\nonumber\\
0&=&-\beta V_1\left(\frac{\partial p}{\partial n}\right)_{\!\!s_1}+\frac{c_P}{T}\left(\frac{\partial T}{\partial n}\right)_{\!\!s_1},
\label{eq:4.4T}
\end{eqnarray}
where we have used eq.~\ref{eq:2.2T} and the definition of $c_P$. Hence,
\begin{equation}
\left(\frac{\partial T}{\partial n}\right)_{\!\!s_1}=V_1\frac{\beta T}{c_P}\left(\frac{\partial p}{\partial n}\right)_{\!\!s_1}.
\label{eq:4.5T}
\end{equation}
Inserting eq.~\ref{eq:4.5T} and eq.~\ref{eq:3.1T} into eq.~\ref{eq:4.2T} and using $nV_1=1$ then yields
\begin{equation}
\left(\frac{\partial p}{\partial n}\right)_{\!\!s_1}=\left(\frac{\partial p}{\partial n}\right)_{\!\!T}+\frac{\beta^2T}{c_P}\left(\frac{\partial p}{\partial n}\right)_{\!\!T}\left(\frac{\partial p}{\partial n}\right)_{\!\!s_1}.
\label{eq:5.1T}
\end{equation}
With eq.~\ref{eq:2.8T}, eq.~\ref{eq:5.1T} takes the simple form
\begin{equation}
\left(\frac{\partial p}{\partial n}\right)_{\!\!s_1}=\left(\frac{\partial p}{\partial n}\right)_{\!\!T}+\frac{c_P-c_V}{c_P}\left(\frac{\partial p}{\partial n}\right)_{\!\!s_1}.
\label{eq:5.2T}
\end{equation}
Using eqs.~\ref{eq:4.1aT}~and~\ref{eq:4.1bT} in eq.~\ref{eq:5.2T}, we obtain the well-known result~\cite{LandauFluids}
\begin{equation}
c_T^2=\frac{c_V}{c_P}\,c_0^2.
\label{eq:5.5T}
\end{equation}
Eqs.~\ref{eq:4.1bT}~and~\ref{eq:5.5T} then determine $\beta^2$ from eq.~\ref{eq:2.8T} in terms of measurable quantities,
\begin{equation}
\beta^2=\frac{c_P}{c_V}\frac{c_P-c_V}{m\,c_0^2\,T}.
\label{eq:5.7T}
\end{equation}

Using the results described above, it is easy to evaluate the pressure change $\delta p$  in terms of the density change $\delta n$ and the change in the entropy per particle $\delta s_1$. Writing the pressure as $p(n,s_1)=p[n,T(n,s_1)]$, we have
\begin{eqnarray}
\delta p&=&\left(\frac{\partial p}{\partial n}\right)_{\!\!s_1}\!\delta n+\left(\frac{\partial p}{\partial s_1}\right)_{\!\!n}\!\delta s_1\nonumber\\
&=&mc_0^2\,\delta n+\left(\frac{\partial p}{\partial T}\right)_{\!\!n}\left(\frac{\partial T}{\partial s_1}\right)_{\!\!n}\!\delta s_1.
\label{eq:6.1T}
\end{eqnarray}
where we have used eq.~\ref{eq:4.1aT}. With eqs.~\ref{eq:3.1T},~\ref{eq:4.1bT},~\ref{eq:5.5T}, and the heat capacity per particle at constant volume  $c_V=T(\partial s_1/\partial T)_n$,  we immediately obtain
\begin{equation}
\delta p=mc_0^2\,\delta n+mc_0^2\,\beta\frac{T}{c_P}\,n\,\delta s_1.
\label{eq:6.3T}
\end{equation}
For the temperature change $\delta T$, we write $T(s_1,n)$, so that
\begin{equation}
\delta T=\left(\frac{\partial T}{\partial s_1}\right)_{\!\!n}\!\delta s_1+\left(\frac{\partial T}{\partial n}\right)_{\!\!s_1}\!\delta n.
\label{eq:6.4T}
\end{equation}
Inserting eqs.~\ref{eq:4.5T}~and~\ref{eq:4.1aT} into eq.~\ref{eq:6.4T} and using $V_1=1/n$, we find
\begin{equation}
\delta T=\frac{T}{c_V}\,\delta s_1+mc_0^2\,\beta\frac{T}{c_P}\frac{\delta n}{n}.
\label{eq:6.5T}
\end{equation}

\subsubsection{Unitary Fermi gas thermodynamics}
For the unitary Fermi gas, universality~\cite{HoUniversalThermo,ThomasUniversal,ThermoLuo} requires that the pressure $p$ and the energy density ${\cal E}$ are functions only of the density and temperature, related by  $p=2\,{\cal E}/3$. Dimensional analysis then shows that the energy density takes the simple form
\begin{equation}
{\cal E}=\frac{3}{5}n\,\epsilon_F(n)\,f_E(\theta),
\label{eq:1.1TP}
\end{equation}
where $\theta\equiv T/T_F$ is the reduced temperature with $T_F$ the local Fermi temperature. For a balanced 50-50 mixture of two spin components of total density $n$, the local Fermi energy is  $k_BT_F=\epsilon_F(n)=mv_F^2/2=\hbar^2(3\pi^2n)^{2/3}/(2m)$. The universal function $f_E(\theta)$ has been measured by Ku et al.,~\cite{KuThermo}, which determines all of the thermodynamic properties. The pressure is then
\begin{equation}
p=\frac{2}{5}n\,\epsilon_F(n)\,f_E(\theta).
\label{eq:1.3TP}
\end{equation}
The entropy density takes a similar form
\begin{equation}
s=nk_B\,f_S(\theta)=n\,s_1(\theta),
\label{eq:1.2TP}
\end{equation}
where $f_S(\theta)$ is determined by  $f_E(\theta)$.

The adiabatic sound speed eq.~\ref{eq:4.1aT} is easily obtained from eq.~\ref{eq:1.3TP}, as eq.~\ref{eq:1.2TP} requires constant $\theta$ for constant $s_1$,
\begin{equation}
c_0^2=\frac{1}{m}\left(\frac{\partial p}{\partial n}\right)_{\!\!\theta}.
\label{eq:1.4TP}
\end{equation}
Eq.~\ref{eq:1.4TP} immediately gives
\begin{equation}
c_0=v_F\sqrt{\frac{f_E(\theta)}{3}}.
\label{eq:1.5TP}
\end{equation}
\begin{figure}[htb]
\begin{center}\
\includegraphics[width=3.0in]{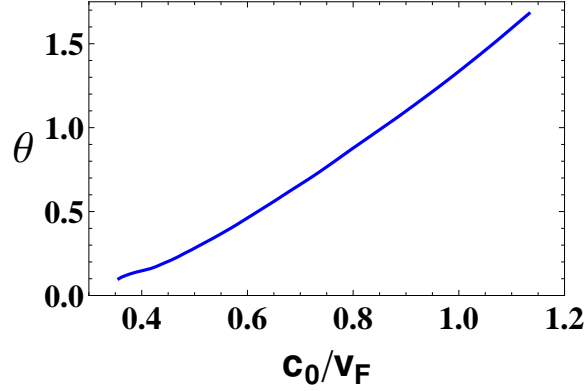}
\end{center}
\caption{Reduced temperature $\theta =T/T_F$ versus adiabatic sound speed $c_0/v_{F}$. For $\theta>0.25$, $\theta$ monotonically increases with $c_0/v_F$, showing that the fitted $c_0/v_F$ can be used as a thermometer to determine $\theta$ in the normal fluid region.
\label{fig:soundspeed}}
\end{figure}
Fig.~\ref{fig:soundspeed} shows how the adiabatic sound speed $c_0$ obtained from the measured equation of state varies with $\theta$.

The heat capacity per particle at constant volume takes a simple form. Using eq.~\ref{eq:1.1TP},
\begin{equation}
c_V=\frac{T}{n}\left(\frac{\partial s}{\partial T}\right)_{\!\!n}=\frac{1}{n}\left(\frac{\partial {\cal E}}{\partial T}\right)_{\!\!n}=k_B\frac{3}{5}f_E'(\theta),
\label{eq:2.1TP}
\end{equation}
where $'$ denotes differentiation with respect to $\theta$.

We determine heat capacity per particle at constant pressure from eq.~\ref{eq:2.8T} by finding $\beta$ from eq.~\ref{eq:3.1T}. Using eq.~\ref{eq:4.1bT}~and~\ref{eq:5.5T}, we have generally
\begin{equation}
\beta =\frac{1}{n} \left(\frac{\partial p}{\partial T}\right)_{\!\!n} \frac{1}{m\, c_T^2}=\frac{c_P}{c_V}\frac{1}{mc_0^2}\frac{1}{n}\left(\frac{\partial p}{\partial T}\right)_{\!\!n}.
\label{eq:beta1}
\end{equation}
For the unitary Fermi gas, where $p=2\,{\cal E}/3$, eq.~\ref{eq:2.1TP} shows that $(\partial p/\partial T)_n=2\,n c_V/3$ and
\begin{equation}
\beta_{UG}=\frac{2}{3}\frac{c_P}{m\,c_0^2}.
\label{eq:beta2}
\end{equation}
Then,  eq.~\ref{eq:2.8T} reduces to
\begin{equation}
c_P-c_V=\frac{4}{9}\frac{T}{m\,c_0^2}\,c_V\,c_P.
\label{eq:beta3}
\end{equation}
With $mc_0^2=2\,k_BT_F\,f_E(\theta)/3$ from eq.~\ref{eq:1.5TP}, eq.~\ref{eq:beta3} takes the simple form
\begin{equation}
\frac{c_P-c_V}{c_V\,c_P}=\frac{1}{k_B}\frac{2}{3}\frac{\theta}{f_E(\theta)}.
\label{eq:beta5}
\end{equation}
Solving eq.~\ref{eq:beta5} for $c_P$, and using $c_V$ from eq.~\ref{eq:2.1TP}, then determines
\begin{equation}
\frac{c_P}{c_V}=\frac{f_E(\theta)}{f_E(\theta)-\frac{2}{5}\theta\,f_E'(\theta)}.
\label{eq:beta4}
\end{equation}
\subsubsection{Determination of the box force}
\label{subsec:boxforce}
We find the force arising from the confining potential along one axis $z$, using the measured density profiles $n_0(z)$. We ignore the variation of the density along the line of site and find $n_0(z)$ from the spatially integrated column density, which is obtained from absorption images. Using eq.~\ref{eq:3.6} with $(\partial p/\partial n)_T =(c_V/c_P)\,m\,c_0^2$ from eq.~\ref{eq:5.5T} and eq.~\ref{eq:4.1bT},
\begin{equation}
\partial_z U_0(z)=-m\,c_0^2[\theta(z)]\,\frac{c_V[\theta(z)]}{c_P[\theta(z)]}\frac{\partial_z n_0(z)}{n_0(z)}.
\label{eq:3.6TP}
\end{equation}
 Here, $c_0$ is given by eq.~\ref{eq:1.5TP} and $c_V/c_P$ is given by eq.~\ref{eq:beta4}. As $\partial_z U_0(z)$ is independent of the scale of $n_0(z)$, we find $\partial_z U_0(z)$ from the the scaled density $\tilde{n}_0(z)=n_0(z)/n_0$, where
$n_0$ is the mean central density. Note that the thermodynamic parameters depend only on the reduced temperature $\theta(z)=\theta_0/[\tilde{n}_0(z)]^{2/3}$, where we determine $\theta_0=T_0/T_F(n_0)$ from the fitted adiabatic sound speed $c_0$.

For the numerical integration, we express $U_0$ in units of $\epsilon_{F0}$, $\tilde{U}_0(z)\equiv U_0(z)/\epsilon_{F0}$ . From eq.~\ref{eq:1.5TP}, we have $mc_0^2(\theta)=mv_F^2 f_E(\theta)/3=2\,\epsilon_F f_E(\theta)/3$. With $\epsilon_F/\epsilon_{F0}=[\tilde{n}_0(z)]^{2/3}$ and eq.~\ref{eq:beta4}, we find
\begin{equation}
\partial_z\tilde{U}_0(z)=-\frac{2}{3}\left[f_E(\theta)-\frac{2}{5}\theta f_E'(\theta)\right]\frac{\partial_z\tilde{n}_0(z)}{[\tilde{n}_0(z)]^{1/3}},
\label{eq:2.10TP}
\end{equation}
which is also obtained from eq.~\ref{eq:3.6} by evaluating $(\partial p/\partial n)_T$ directly from eq.~\ref{eq:1.3TP}.
\begin{figure}[htb]
\begin{center}\
\includegraphics[width=3.5in]{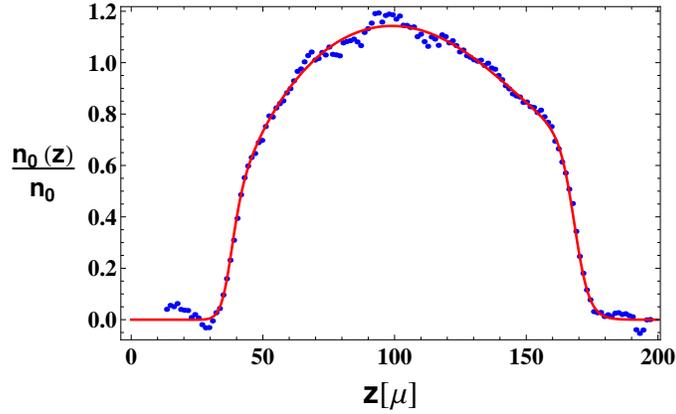}
\end{center}
\caption{Measured background density $n_0(z)$ (blue dots) versus $z$ in $\mu$m, showing the fit of eq.~\ref{eq:fit} (red curve).
\label{fig:backgroundfit}}
\end{figure}
To evaluate the derivative $\partial_z\tilde{n}_0(z)$ on the right hand side of eq.~\ref{eq:2.10TP}, we first fit $\tilde{n}_0(z)=n_0(z)/n_0$ with an analytic function,
\begin{equation}
f(z)= \frac{\tanh[(z - z_{10})/w_1] -
    \tanh[(z - z_{20})/w_2]}{2}\,\sum_n a_n\,z^n.
\label{eq:fit}
\end{equation}
The difference of the $\tanh$ functions produces a top-hat shape of nominal width $z_{20}-z_{10}$ and slopes on each side determined by $w_1$ and $w_2$. The flat top is modulated by the multiplying polynomial.  Fig.~\ref{fig:backgroundfit} shows a typical fit using a fifth order polynomial.

\begin{figure}[htb]
\begin{center}\
\includegraphics[width=3.0in]{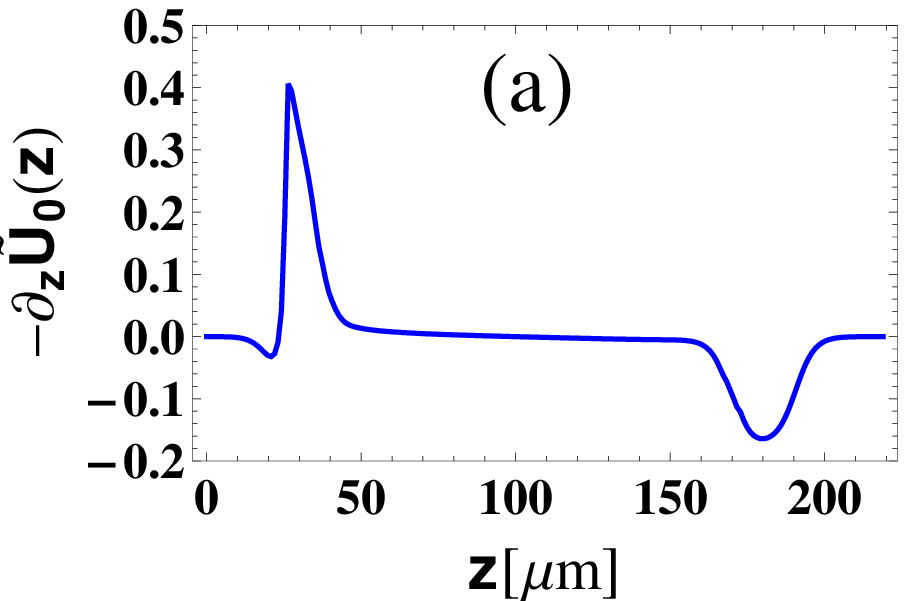}\hspace{0.25in}\includegraphics[width=3.0in]{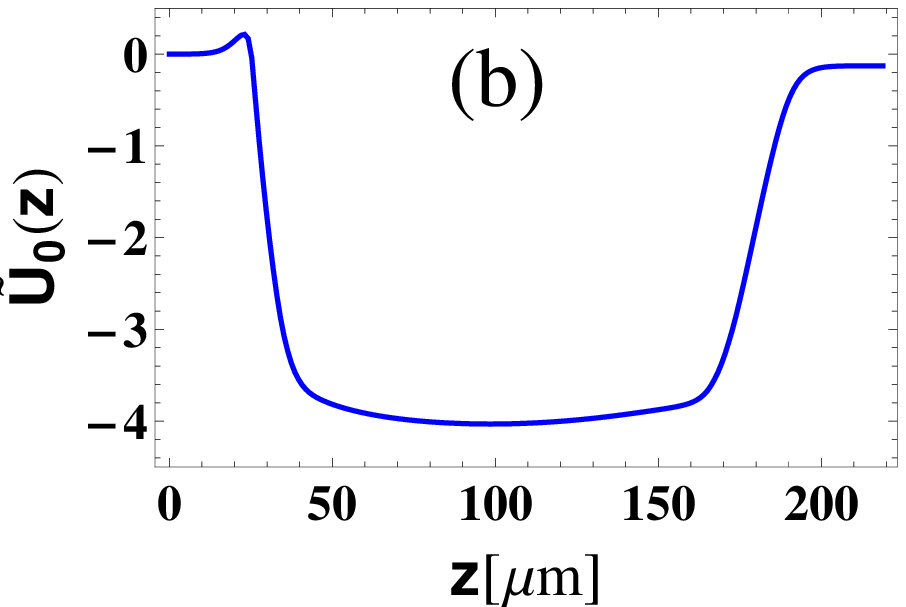}
\end{center}
\caption{Box force profile (a) determined from the measured background density $n_0(z)$ using eq.~\ref{eq:2.10TP}; (b) Corresponding box potential energy in units of $\epsilon_{F0}$ obtained by integrating the force.  Note that the curvature at the bottom of the box potential energy arises from curvature in the bias magnetic field, which produces a confining harmonic potential.\label{fig:force}}
\end{figure}

Eq.~\ref{eq:2.10TP} then yields the force profile, $-\partial_z\tilde{U}_0(z)$, Fig.~\ref{fig:force}(a).  The corresponding box potential is then $\tilde{U}_0(z)=\int_0^z dz'\,\partial_{z'}\tilde{U}_0(z')$,  Fig.~\ref{fig:force}(b).

\subsubsection{Hydrodynamic relaxation time}

The local shear viscosity can be written as $\eta =\tau p$~\cite{BluhmSchaeferModIndep,BluhmSchaeferLocalViscosity}, where $p$ is the pressure and $\tau$ is the hydrodynamic relaxation time. We estimate $\tau$ for the unitary Fermi gas using eq.~\ref{eq:1.3TP} for $p$ and
\begin{equation}
\eta=\hbar\,n\,(\alpha_0\,\theta^{3/2}+\alpha_2),
\label{eq:viscosity1}
\end{equation}
with $n$ the local density, $\alpha_0=2.77$ and $\alpha_2=0.25$~\cite{BluhmSchaeferLocalViscosity}, as discussed in the main paper. Then with $\tau=\eta/p$,
\begin{equation}
\tau=\frac{5}{2}\tau_F\frac{\alpha_0\,\theta^{3/2}+\alpha_2}{f_E(\theta)},
\label{eq:relaxtime}
\end{equation}
where $\tau_F=\hbar/\epsilon_F$ is the Fermi time. For our experiments, where $\epsilon_F\simeq k_B\times 0.15\,\mu$K and $\theta\simeq 0.5$, we find $\tau=0.13$ ms, which is small compared to the period at the highest frequency of 800 Hz that is employed in the experiments.


\begin{thebibliography}{24}
\bibitem [{Tem()}]{TempGrad}%
  \BibitemOpen
  \href@noop {} {}\bibinfo {note} {In zeroth order, where $\eta\rightarrow 0$,
  the cloud expands adiabatically and the spatial profile obeys a scale
  transformation. Then, as the cloud expands, the local density $n$ and hence
  the temperature $T\propto n^{2/3}$, decrease by a spatially uniform
  time-dependent scale factor, so that the initially uniform temperature
  profile remains
  uniform~\cite{CaoViscosity,JosephShearNearSF,BluhmSchaeferLocalViscosity}.}\BibitemShut
  {Stop}%
\end{thebibliography}
\end{document}